\documentclass[preprint]{aastex}
\usepackage{graphicx,natbib,amsmath,apjfonts}

\newcommand{\bbf}{}

\slugcomment{Accepted by ApJ (2014)}

\shorttitle{HDO and H$_2$O in the solar nebula}
\shortauthors{Albertsson et al.}

\begin{document}

\title{Chemo-dynamical deuterium fractionation in the early solar nebula: \\ 
The origin of water on Earth and in asteroids and comets.}

\author{T. Albertsson, D. Semenov and Th. Henning}
\affil{Max-Planck-Institut f\"ur Astronomie, K\"onigstuhl 17, 69117 Heidelberg, Germany}
\email{albertsson@mpia.de}

\begin{abstract}
Formation and evolution of water in the Solar System and the origin of water on Earth constitute one of the most interesting questions in astronomy. The prevailing hypothesis for the origin of water on Earth is by delivery through water-rich small Solar system bodies. In this paper, the isotopic and chemical evolution of water during the early history of the solar nebula, before the onset of planetesimal formation, is studied. A gas-grain chemical model that includes multiply-deuterated species and nuclear spin-states is combined with a steady-state solar nebula model. To calculate initial abundances, we simulated 1~Myr of evolution of a cold and dark TMC1-like prestellar core. Two time-dependent chemical models of the solar nebula are calculated over 1~Myr: (1) a laminar model and (2) a model with 2D turbulent mixing. 
We find that the radial outward increase of the H$_2$O D/H ratio is shallower in the chemo-dynamical nebular model compared to the laminar model. This is related to more efficient de-fractionation of HDO via rapid gas-phase processes, as the 2D mixing model allows the water ice to be transported either inward and thermally evaporated or upward and photodesorbed. The laminar model shows the Earth water D/H ratio at $r~\la~2.5$~AU, while for the 2D chemo-dynamical model this zone is larger, $r~\la~9$~AU. Similarly, the water D/H ratios representative of the Oort-family comets, $\sim~2.5-10\times~10^{-4}$, are achieved within $\sim~2-6$~AU and $\sim~2-20$~AU in the laminar and the 2D model, respectively. We find that with regards to the water isotopic composition and the origin of the comets, the mixing model seems to be favored over the laminar model. 
\end{abstract}

\section{Introduction}
\label{sec:intro}
Water is one of the most abundant molecules detected in interstellar space, \bbf{found throughout all stages of the stellar 
evolution, from prestellar cores till protoplanetary disks}. The ultimate role of liquid water for the origin of life on Earth is 
well recognized in biochemistry and astronomy \citep[e.g., ][]{1993Icar..101..108K, 2006BGD.....3...23H, 2009..BioChem..284, 
2010A&ARv..18..383J, 2013ApJ...774L...3L}. One of the main difficulties in tracing the origin of the Earth water is strong 
evidence that it came from exogenous sources, since the solar nebula was too hot in the terrestrial planet-formation zone to form 
water-rich planetesimals from smaller icy precursors 
\cite[][]{1998AREPS..26...53B,2000M&PS...35.1309M,2011Icar..212..416M,2012Icar..221..859E}. In addition, no evidence for hydrous 
silicates in the inner regions of protoplanetary disks has been found \citep{2010ApJ...721..431J}. 

The most promising mechanism of water delivery to Earth is bombardment by carbonaceous asteroids and/or comets that have formed in 
the ice-rich, outer region of the young Solar System \citep[e.g.][]{1961Natur.190..389O, 1992Natur.358...43O, 2000M&PS...35.1309M, 
2011NatGe...4...74R, 2013ApJ...767...54I}. This idea is supported by modern sophisticated dynamical models of the early Solar 
System, which show that complex gravitational interactions between Jupiter and Saturn had likely scattered other small bodies into 
the inner and far outer regions \citep[see e.g.][]{2005Natur.435..462M, 2005Natur.435..466G, 2005Natur.435..459T, 
2012M&PS...47.1941W}. \bbf{More evidence of the Solar System's dynamical past comes from} the presence of crystalline silicates 
annealed at temperatures above 800~K in cometary samples collected by the {\it Stardust} mission \citep{2012M&PS...47..660W} and 
revealed by infrared observations of comets \citep[e.g.][]{Crovisier1997, 2006ApJ...651.1256K} suggests that transport or highly 
energetic local processes occurred in the solar nebula. The importance of dynamical activity of the young solar nebula and 
protoplanetary disks on its chemical and isotopic evolution was also demonstrated in numerous theoretical studies 
\citep[e.g.][]{1984ApJ...287..371M, Cyr_ea98,1999ApJ...526..314A, G01, 2004A&A...415..643I, 2004ApJ...616.1265B, 
2006ApJ...644.1202W, 2007A&A...463..369T, 2007ApJ...660..441W, 2009ApJ...703..479W, 2011ApJS..196...25S, Heinzeller_ea11, 
2012M&PS...47..660W}. 

The individual contributions from particular exogenous sources is still heavily debated. Water has likely formed on surfaces of 
dust grains even before the formation of the solar nebula in the protosolar molecular cloud and inherited specific isotopic 
composition and ortho/para ratios from that epoch \citep[e.g.,][]{2010ApJ...725L.172J, 2011ApJ...741L..34C,2013A&A...550A.127T}, 
which may have been preserved in primitive Solar System bodies such as comets. It is plausible to assume though that the water 
isotopic composition (and may be even the ortho/para ratio) could have been partly or fully reset during the violent physical 
conditions in the inner, dynamically active solar nebula \citep[e.g.,][]{2000SSRv...92..201R, 2004ApJ...616.1265B}. Other 
processes such as grain growth and sedimentation have shown to bear a profound effect on the water abundance and deuterium 
fractionation \citep{2011ApJ...726...29F,2011ApJ...727...76V,ANDES}. The importance of the ortho- and para-states of H$_{2}$ for 
deuterium 
fractionation is well established and also has to be taken into account \citep[e.g.,][]{2006A&A...449..621F}.

\bbf{The water D/H ratio remains the most essential probe for disentangling the individual contributions from the various 
exogenous sources.} Since comets are reservoirs of pristine material in the solar nebula, a comparison of the water D/H ratios 
measured in comets and asteroids to that measured in protoplanetary disks is vital. These observations are challenging because 
most of the water is frozen out onto dust grains in disks around cool Sun-like stars \citep{Dominik_ea05a, 2010A&A...521L..33B, 
2011Sci...334..338H}. Furthermore, disks have relatively compact sizes ($\la 100-1000$~AU) and are not massive ($\la 
0.01M_\odot$). Therefore, detailed high-resolution studies remain a challenge. Ground-based observations are further compromised 
by the absorption and veiling of water vapor lines in the Earth atmosphere. \bbf{In order} to better understand the origin of 
water on Earth, other Solar System bodies, and possibly in other exoplanetary systems, one has to study the chemical evolution of 
water prior 
and during the onset of planet formation.

\bbf{The water evolution begins in prestellar cores, where the low temperatures cause many molecules to freeze-out onto grains, 
and the gas remains largely void of molecules \citep{2010A&A...521L..29C}. This drives a rich surface-chemistry along with an 
efficient deuterium fractionation, resulting in high water abundances on the grains ($\sim 10^{-4}$ relative to H) and D/H ratios 
($\sim 10^{-2}$). As the cores contracts, and begins to heat up, volatiles are released into the gas. Water has been detected 
towards several class 0 protostellar objects \citep[e.g.][]{1999A&A...342L..21C, 2010A&A...521L..30K, 2013ApJ...769...19V, 
2013arXiv1308.5119M}, but only a few  detections of HDO (and D$_{2}$O) have allowed determination of D/H ratios, with values$\sim 
10^{-4}$ in the hot cores, and retaining $\sim 10^{-2}$ in the cold outer regions \citep{2010ApJ...725L.172J, 2011A&A...527A..19L, 
2012A&A...539A.132C, 2013A&A...549L...3P, 2013ApJ...768L..29T, 2013A&A...553A..75C}. }

 
Several detections \bbf{of water} have been done in protoplanetary disks \citep[e.g.][]{2005ApJ...622..463P, 2010A&A...518L.124M, 
2010A&A...521L..33B, 2012A&A...544L...9F}, \bbf{including also several using \textit{PACS/Herschel} \citep{2011Sci...334..338H, 
2012A&A...538A..20T, 2012A&A...544L...9F, 2012A&A...540A..84H, 2012A&A...538L...3R}} within the WISH program \citep[Water In 
Star-forming Regions with Herschel][]{2011PASP..123..138V}. \bbf{In particular,} hot water vapor ($\ga 500$~K) has been discovered 
in the inner, planet-forming regions of protoplanetary disks around young solar-mass T~Tauri stars, mainly, by infrared 
spectroscopy with the \textit{Spitzer} satellite 
\citep{2008Sci...319.1504C,Pontoppidan_ea08,2010ApJ...720..887P,2011ApJ...743..147N,2011ApJ...734...27T,2012ApJ...745...90B}. 

The first detection of the ground transition of HDO in absorption in the disk around DM~Tau was reported by 
\citet{2005ApJ...631L..81C}. Although they have not observed H$_2$O, \citet{2005ApJ...631L..81C} predicted water D/H ratios of $> 
0.01$, based on H$_{2}$O measurements in molecular clouds and protostellar envelopes. This discovery was later disputed by 
\citet{2006A&A...448L...5G}. \citet{2008ApJ...681.1396Q} have derived upper limits for the water column densities in the disk 
around TW~Hya, $\la 9.0\times10^{14}$~cm$^{-2}$, but were not able to determine its D/H ratio. 

The discovery of the ground-state rotational emission of both spin isomers of H$_2$O from the outer, $\ga 50-100$~AU disk around 
TW~Hya with \textit{Herschel} allowed to derive a water abundance of $\approx 10^{-7}$ relative to H$_2$ 
\citep{2011Sci...334..338H}. The authors also derived a water ortho/para ratio of $0.77 \pm 0.07$, which indicates that the water 
formation took place on cold dust grains with temperatures $\sim 10-20$~K. In contrast, the water ortho/para ratios measured in 
comets are indicative of higher temperatures of $\sim 25-50$~K, more typical for the inner solar nebula at $\la 20-30$~AU 
\citep{Mumma_ea87, Woodward_ea07, BM_ea09, Shinnaka_ea12, Bonev_ea13a}. 

Studies of carbonaceous chondrites have revealed that their phyllosilicates have D/H ratios that are very similar to that in the 
Earth's ocean water, $\sim~1.56~\times$~10$^{-4}$, with as large overall water content as $\la 20\%$, thus favoring asteroids as 
the main water source \citep[e.g.,][]{1992Natur.358...43O,2006BGD.....3...23H,2012E&PSL.313...56M,2012Sci...337..721A}. Since 
observations of Oort-family comets have yielded water D/H ratios in their comae of about 2-3 times higher that in the Earth' 
oceans \citep[][and references therein]{2000SSRv...92..201R}, and because of the low amount of volatile noble gases on Earth 
\citep[e.g., ][]{Marty01012013}, it was concluded that comets contributed at most up to $\sim 10\%$ to the water delivery on 
Earth. However, recent HDO observation of the short-period, Jupiter-family comet Hartley-2 with {\it Herschel} by 
\citet{2011Natur.478..218H} revealed a water D/H ratio of $1.6\times 10^{-4}$, which is very close to the Earth value. 
Additionally, \citet{2013ApJ...774L...3L} studied the Jupiter-family comet 45P/Honda-Mrkos-Pajdusakova but could not detect HDO 
with an upper water D/H ratio $2 \times 10^{-4}$. Their upper limit is still consistent with the value found by 
\citet{2011Natur.478..218H} and aids in confirming the diversity of D/H ratios between different comet populations. These 
observations have reignited again debates about the relative role of asteroids and comets in the origin of the Earth water. One 
has to bear in mind, however, that a major problem could be that the gases in cometary comae may have lower D/H ratios compared to 
the D/H ratios of the nucleus, as indicated by experiments of \citet{2012P&SS...60..166B}. 

\bbf{While we have made many advances in our understanding in regards of the origin of Earth's oceans, we still do not fully 
understand the reprocessing of water during the protoplanetary disk stage, nor is there any consensus for contribution from the 
various exogenous sources. }
In this paper, we focus on the history and evolution of the early solar nebula and investigate the chemistry of frozen and gaseous 
water at the stage when planetesimals have not yet formed, and the micron-sizes dust grains were the dominant population of 
solids. We use an extended gas-grain chemical model that includes multiply-deuterated species, high-temperature reactions, surface 
reactions, and nuclear spin-states processes. This network is combined with a 1+1D steady-state $\alpha$-viscosity nebula model to 
calculate molecular abundances and D/H ratios within the first 1~Myr of its lifetime. We consider both the laminar case with 
chemistry not affected by dynamics and a model with 2D turbulent mixing transport. The organization of the paper is the following. 
In Section~\ref{sec:model} we describe our physical and chemical model of the solar nebula and introduce the considered deuterium 
processes. The results are presented in Sect.~\ref{sec:results}, followed by Discussion and Conclusions.

\section{Model}
\label{sec:model}
\subsection{Physical model}
\label{sec:model:physics}
\begin{figure*}[!htb]
\centering
\includegraphics[width=0.32\textwidth,angle=90]{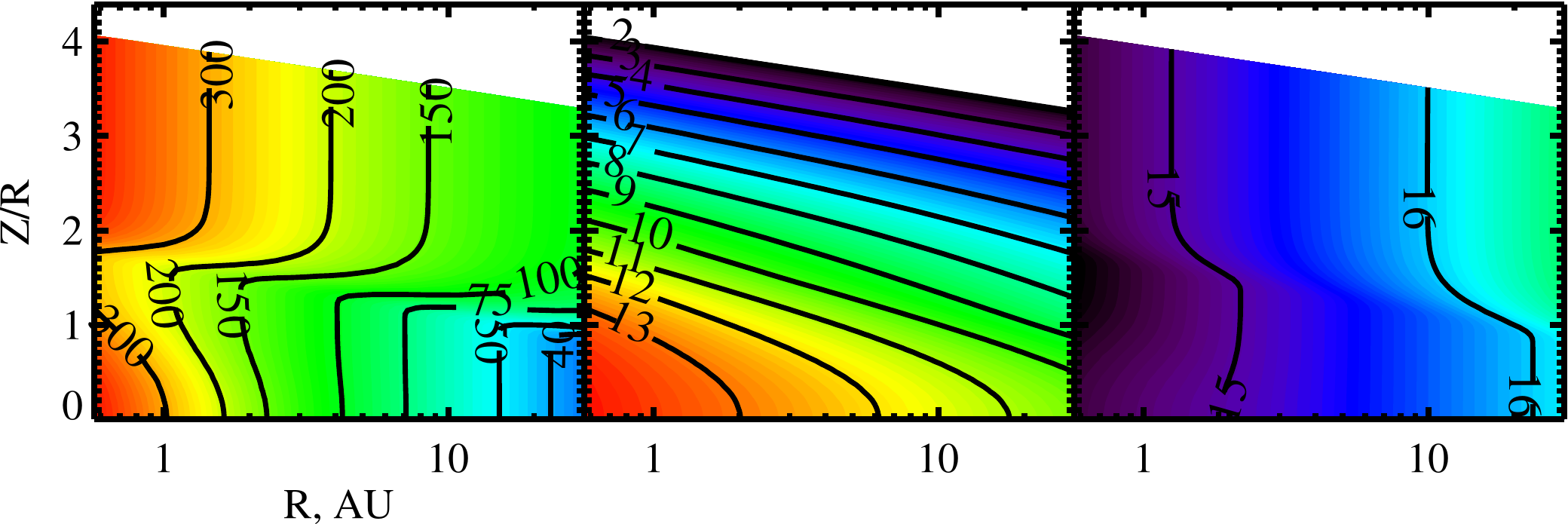}
\caption{(Left to right) Distributions of temperature in K, 
density in cm$^{-3}$ ($\log_{10}$ scale), and diffusion coefficient in cm$^2$\,s$^{-1}$
($\log_{10}$ scale) in the solar nebula model. }
\label{fig:disk_struc}
\end{figure*}

We used a similar disk structure as in \citet{2011ApJS..196...25S}, but adopted parameters characteristic of the young Sun and 
early solar nebula. The physical model is based on a 1+1D steady-state $\alpha$-model of a flaring protoplanetary disk described 
by \citet{1999ApJ...527..893D}, see Fig.~\ref{fig:disk_struc}. We used the parametrization of \citet{1973A&A....24..337S} of the 
turbulent viscosity $\nu$ in terms of the characteristic scale height $H(r)$, the sound speed $c_{\rm s}(r,z)$, and the 
dimensionless parameter $\alpha$:
\begin{equation}
 \nu(r,z) = \alpha\,c_{\rm s}(r,z)\,H(r).
 \end{equation}
From observations and detailed MHD modeling the $\alpha$ parameter has values $\sim 0.001-0.1$ 
\citep[][]{Andrews_Williams07,Guilloteau_ea11a,2011ApJ...735..122F}. A constant value of $\alpha=0.01$ was used in our 
simulations. Equal gas and dust temperatures were assumed.

The central star has a mass of $1M_\odot$, a radius of $1.2R_\odot$, and an effective temperature of 6\,000~K. The non-thermal FUV 
radiation from the young Sun is represented by the interstellar radiation field of \citet{G}, with the un-attenuated intensity at 
100~AU of $\chi_*(100)=10\,000$ $\chi_{\rm draine}$. For the X-ray luminosity of the young Sun we adopted a typical value for 
T~Tauri stars, $10^{30}$~erg\,s$^{-1}$. The model disk has an inner radius of $0.03$~AU (dust sublimation front, $T\approx 
1\,600$~K), an outer radius of $800$~AU, and an accretion rate of $\dot{M}= 10^{-8}\,M_\odot$\,yr$^{-1}$. The mass of the solar 
nebula disk is not well-known but studies suggest it to be $\gtrsim$ 0.1 M$_{\odot}$ \citep[see e.g.][]{1993LPI....24.1225R, 
2007ApJ...671..878D}, and hence we adopt a disk model with a total mass of $M=0.11\,M_\odot$. 

We assume that the dust grains are uniform spherical particles, with a radius of $a_{\rm d}=0.1\,\mu$m, made of amorphous 
silicates with olivine stoichiometry, with a density $\rho_d=3$~g\,cm$^{-3}$ and a dust-to-gas mass ratio $m_{d/g}=0.01$. The 
surface density of sites is $N_s=1.5\times 10^{15}$~sites\,cm$^{-2}$, and the total number of sites per each grain is 
$S=1.885\times 10^6$. No substantial grain growth is assumed in this early disk phase, an assumption which may be challenged in 
other studies. 

In our simulations we assume that gas-phase species and dust grains are well mixed and coupled, and transported with the same 
diffusion coefficient
\begin{equation}
D_{\rm turb}(r,z) = \nu(r,z)/Sc,
\end{equation}
where $Sc=1$ is the Schmidt number that describes the efficiency of turbulent diffusivity \citep[see e.g.][]{1973A&A....24..337S, 
2004ApJ...614..960S}. We treat diffusion of mantle materials similarly to gas-phase molecules, without relating it to individual 
grain dynamics. 

As boundary conditions for mixing, we assume that there is no inward and outward diffusion across boundaries of the solar nebula, 
and that there is no flux through its central plane. All the equations are solved on a non-uniform staggered grid consisting of 65 
radial points (from 0.5 to 800~AU) and 91 vertical points. 

The physical structure of the solar nebula and the chemical model with and without turbulent mixing are used to solve chemical 
kinetics equations \citep[see Eq.~3 in][]{2011ApJS..196...25S}. The equations of chemical kinetics are integrated together with 
the diffusion terms in the Eulerian description, using a fully implicit 2D integration scheme, and a sparse matrix formalism for 
inversion of the Jacobi matrices \citep{Semenov_ea10}.

\subsection{Chemical network and model}
\label{sec:model:chemistry}
To calculate photoreaction rates through the various disk environments, we adopt precomputed fits of \citet{2006FaDi..133..231V} 
for a 1D plane-parallel slab, using the Draine far-UV interstellar radiation field. The self-shielding of H$_{2}$ from 
photodissociation is calculated by Equation (37) from \citet{1996ApJ...468..269D} and the shielding of CO by dust grains, H$_{2}$, 
and its self-shielding are calculated using the precomputed table of \citet{1996A&A...311..690L}. We do not take the Ly$_\alpha$ 
radiation into account.

We used the gas-grain chemical network developed by \citet{2013ApJS..207...27A}, which includes an extended list of fractionation 
reactions for up to triply-deuterated species. In addition, the high-temperature reaction network from \citet{2010ApJ...721.1570H, 
2012ApJ...756..104H} has been added. Recently, it was realized that the ortho/para ratio of H$_{2}$ and other species can lower 
the pace of deuterium fractionation \citep{2006A&A...449..621F, 2009A&A...494..623P}. The internal energy of ortho-H$_2$ is higher 
than that of para-H$_2$, which helps to overcome the barrier of the backward reaction of deuterium enrichment. Consequently, it 
results in a lower degree of deuterium fractionation in a medium having a sufficient amount of ortho-H$_{2}$ 
\citep{2006A&A...449..621F}. Given the importance of the ortho/para ratios of H$_2$ and H$_3^+$ for efficiency of deuterium 
fractionation, particularly, in the solar nebula regions with $T\ga 15-30$~K, the nuclear spin-state processes and ortho-para 
states of H$_{2}
$, H$_{2}^{+}$ and H$_{3}^{+}$ were added to our chemical network. 

\bbf{Reaction rates for a small number of reactions have already been measured or theoretically predicted. For this, we have added 
rates from several sources \citep{1990JChPh..92.2377G, 2004A&A...418.1035W, 2004A&A...427..887F, 2009A&A...494..623P, 
2011PhRvL.107b3201H}, including reaction rates for the H$_{3}^{+}$ + H$_{2}$ system by \citet{2009JChPh.130p4302H}. In order to 
include the nuclear spin states of any reactions including H$_{2}$, H$_{2}^{+}$, H$_{3}^{+}$, or any of their isotopologues, we 
employed a separation scheme similar to that described in \citet{2013A&A...554A..92S}. This routine will be adopted and evaluated 
in Albertsson et al. 2014, Submitted.} Contrary to \citet{2013A&A...554A..92S}, we allow reactions without H$_3^+$ or H$_2^+$ as 
reactants to form both ortho- and para-H$_2$, since we are interested in modeling chemistry in the inner warm nebula region. The 
size of the original network was reduced to facilitate the performance of our 2D chemo-dynamical model by allowing 
fractionation only for species with $<4$ hydrogen atoms, $<4$ carbon atoms, and species not larger than $7$ atoms.

We find that the inclusion of the high-temperature reactions can increase the abundance of HDO and H$_2$O by up to a factor of 2 
inside the snow line ($\lesssim 2-3$ AU). In contrast, the reduction of the original network to $\sim 39\,000$ reactions and $\sim 
1\,300$ species bears only insignificant effect on the calculated time-dependent abundances of the H$_2$O isotopologues. 

\begin{deluxetable}{lclc}
\centering
\tabletypesize{\small}
\tablecaption{Initial abundances for the models (fractional abundances). \label{tab:IA}}
\tablehead{
\colhead{Species}		&	\colhead{Abundances}	& \colhead{Species}		&	\colhead{Abundances}
}
\startdata
p-H$_{2}$				&	4.783 $\times 10^{-1}$	&	CH$_{4}$ (ice)			&	5.832 
$\times 10^{-6}$	\\
He					&	9.750 $\times 10^{-2}$	&	N$_{2}$				&	5.135 
$\times 10^{-6}$	\\
o-H$_{2}$				&	2.146 $\times 10^{-2}$	&	O$_{2}$ (ice)			&	4.619 
$\times 10^{-6}$	\\
H					&	2.681 $\times 10^{-4}$	&	O					&	
2.661 $\times 10^{-6}$	\\
H$_{2}$O (ice)			&	7.658 $\times 10^{-5}$	&	N$_{2}$ (ice)			&	2.464 $\times 
10^{-6}$	\\
CO (ice)				&	5.219 $\times 10^{-5}$	&	C$_{3}$H$_{2}$ (ice)	&	6.493 $\times 
10^{-6}$	\\
CO					&	1.816 $\times 10^{-5}$	&	HD					&	
1.479 $\times 10^{-5}$	\\
HNO (ice)				&	3.541 $\times 10^{-6}$	&	O$_{2}$				&	9.645 
$\times 10^{-6}$	\\
HDO (ice)				&	2.887 $\times 10^{-6}$	&	NH$_{3}$ (ice)			&	8.663 
$\times 10^{-6}$	\\
D					&	2.731 $\times 10^{-6}$	&	OH					&	
2.295 $\times 10^{-6}$	
\enddata
\end{deluxetable}
We also added neutral-neutral reactions. The pre-exponential rate factors were calculated using a simple collisional theory and 
the reactions barriers are adopted from \citet{1994GeCoA..58.2927L}. The added neutral-neutral reactions are listed below: 
\begin{eqnarray}
\small
\begin{tabular}{lclc}
HDO + $p$-H$_{2}$ 	&$\Rightarrow$&	H$_{2}$O + HD;		&	2.0$\times 10^{-10}$ cm$^{-3}$ s$^{-1}$,	5170 K	
\nonumber\\
HDO + $o$-H$_{2}$ 	&$\Rightarrow$&	H$_{2}$O + HD;		&	2.0$\times 10^{-10}$ cm$^{-3}$ s$^{-1}$,	5000 K	
\nonumber\\
H$_{2}$O + HD		&$\Rightarrow$&	HDO + $p$-H$_{2}$;	&	5.0$\times 10^{-11}$ cm$^{-3}$ s$^{-1}$,	4840 K	
\nonumber\\
H$_{2}$O + HD		&$\Rightarrow$&	HDO + $o$-H$_{2}$;	&	1.5$\times 10^{-10}$ cm$^{-3}$ s$^{-1}$,	4840 K	
\nonumber
\end{tabular}
\label{eq:neutneut}
\end{eqnarray}
\bbf{However, we find that the addition of these reactions has no significant effect on our results.} For initial abundances we 
model the chemical evolution of a cold and dark TMC 1-like prestellar core, \bbf{temperature 10 K, density $10^{4}$ cm$^{-3}$ and 
extinction $A_{V} = 10$ mag}, for 1~Myr using the ``low metals'' initial abundance set from 
\citet[][Table~11]{1996A&A...311..690L}, \bbf{with an initial H$_{2}$ ortho:para ratio 1:100 (as H$_{2}$ exist in para-form in 
cold environments)}. We note that the initial H$_{2}$ ortho:para plays an essential role here, as it affects the deuterium 
chemistry, and adopting the statistical 3:1 ortho:para for H$_{2}$ as initial abundance for our TMC 1 model, ice HDO abundances 
drop by a factor of 10, and gas HDO abundances by a factor 20. Gas and ice H$_{2}$O are not significantly affected. The most 
abundant species in the initial abundances are listed in Table~\ref{tab:IA}. 

\section{Results}
\label{sec:results}
Our main set of chemical simulations consists of the two runs: (1) the laminar nebular model without transport processes, and (2) 
the 2D-mixing model with $Sc=1$. The water abundance is limited by the initial oxygen abundance in the models and change 
throughout the disk. Given the rapid, $\sim 2-3$~Myr, \bbf{dynamical} evolution of solids in the nebula, we modeled chemistry only 
within 1~Myr. 

\begin{figure*}[!htb]
\centering
\includegraphics[width=1.00\textwidth]{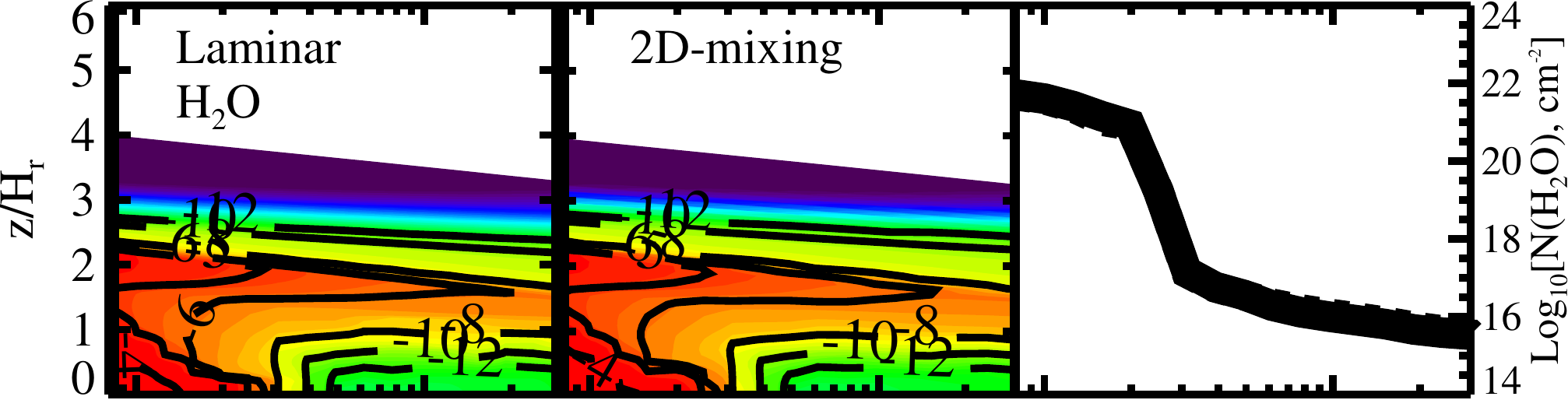} \\
\includegraphics[width=1.00\textwidth]{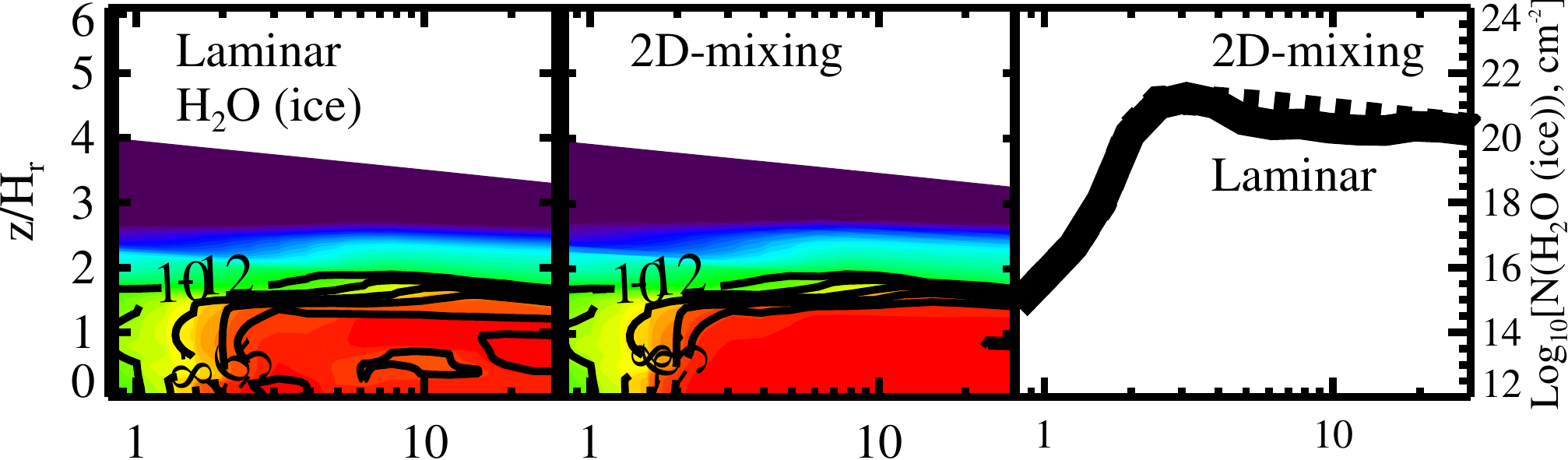} 
\caption{The distributions of gaseous (upper row) and solid (bottom row) water abundances (wrt total H) in the solar nebula 
between 0.8 and 30~AU at 1~Myr. The laminar model is shown on the left panel, the 2D-mixing model is shown in the middle panel. 
The vertically integrated column densities are compared in the right panel, with the laminar model depicted by solid line and the 
2D-mixing model depicted by dashed lines. The thickness of these lines correspond to intrinsic uncertainties in the calculated 
abundances and thus column densities. }
\label{fig:H2O}
\end{figure*}

\begin{figure*}[!htb]
\centering
\includegraphics[width=1.00\textwidth]{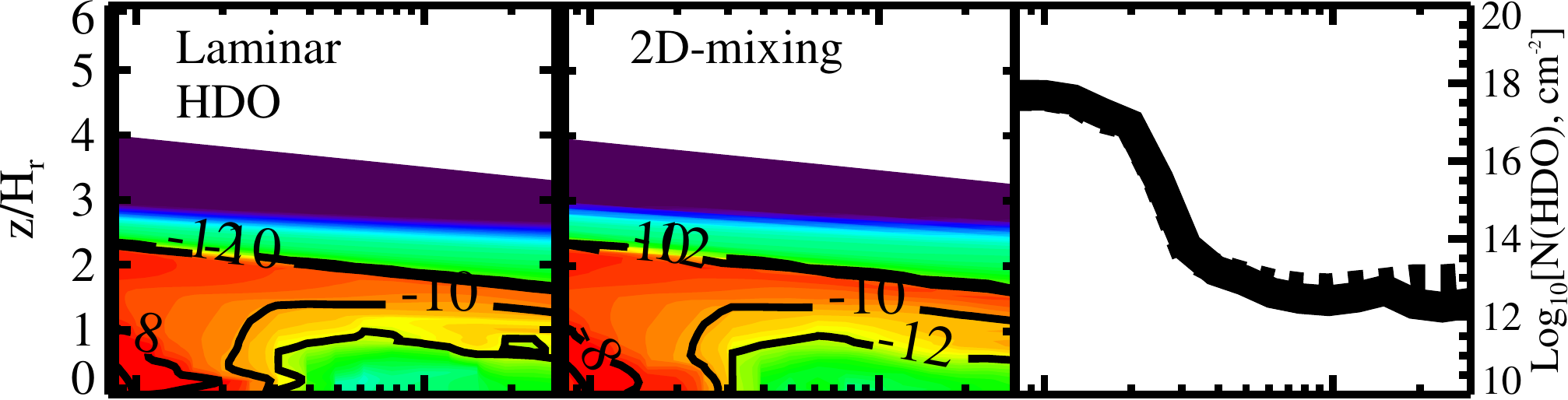} \\
\includegraphics[width=1.00\textwidth]{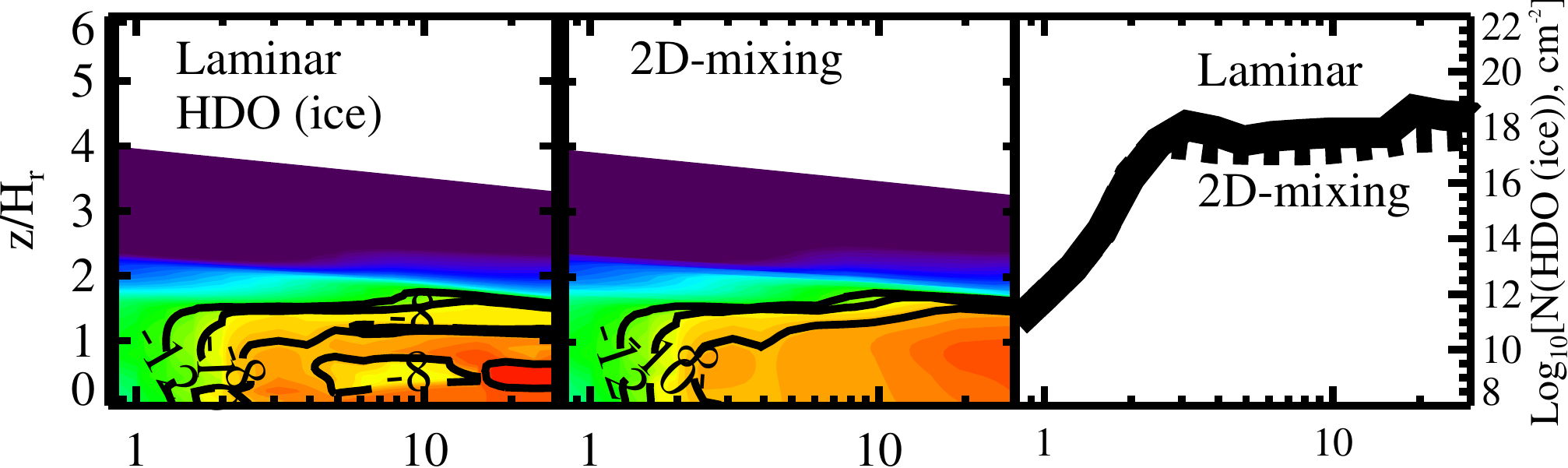} 
\caption{The same as Fig.~\ref{fig:H2O} but for HDO.}
\label{fig:HDO}
\end{figure*}

In Figure~\ref{fig:H2O} we show the gas- and solid-phase water abundances between 0.8 and 30 AU for the laminar (left panel) and 
the 2D-mixing model (middle panel). \bbf{D/H ratios from column densities are shown in the right panel}. Column densities vary 
strongly with scale height. Both the laminar and 2D-mixing show similar relative abundances of water (with respect to H) vapor 
going from $10^{-5}$ close to the midplane and quickly dropping to $< 10^{-12}$ at around 0.15 scale heights in the inner regions, 
while in the outer regions relative abundances in the cold midplane are $\sim 10^{-12}$, increasing to $\sim 10^{-6}$ at 0.15 
scale heights and then dropping again to $< 10^{-12}$ at $\sim$ 0.4 scale heights. Also for water ice the two models show similar 
the relative abundances, with only smaller differences arising in the outer region $\sim$ 10 AU. In the warm inner regions 
relative abundances are $\sim 10^{-8} - 10^{-6}$ in the midplane, dropping $< 10^{-12}$ at $\sim$ 0.15 scale heights. Further 
out in the disk, beyond the snow line at $\sim 2-3$ AU, a thick layer of abundant ices with relative abundances $\sim 10^{-4}$ 
stretches between 0.15 up to 0.3 scale heights, after which we see again a quick drop with relative abundances going below 
$10^{-12}$. The calculated vertical column densities for both models are compared in the right panel.

The same panels are shown in Figure~\ref{fig:HDO} for HDO. Here layers of different relative abundances are more clear, with 
relative abundances of HDO vapor in the inner disk of $\sim 10^{-10}$, dropping quickly $<10^{-12}$ at 0.15 scale heights. In the 
outer regions we have a low vapor abundance in the midplane of $\sim 10^{-14} - 10^{-12}$, increasing up to $10^{-10}$ in the 
molecular layer and then at around 0.2 scale heights dropping $< 10^{-12}$. For HDO ice both models show relative abundances $\sim 
10^{-12}$ in the inner region and a thick layer of abundances $\sim 10^{-8}$ between 0.15 - 0.3 scale heights in the outer 
regions. While the 2D-mixing model show a smooth decline in abundances towards the inner disk, the laminar model show larger 
variations, especially evident at 5$-$30 AU. As can be clearly seen, the snow line is not affected by slow diffusive transport and 
is located at $\sim 2-3$ AU in both models, thus leaving dust grains in the Earth-formation zone barren of ices. Hence, the water 
delivery to Earth could occur only at a later stage, when dust grains were assembled to larger planetesimals that were able to 
reach 1~AU without complete loss of volatile materials.

\begin{figure}[!htb]
\centering
\includegraphics[width=0.48\textwidth]{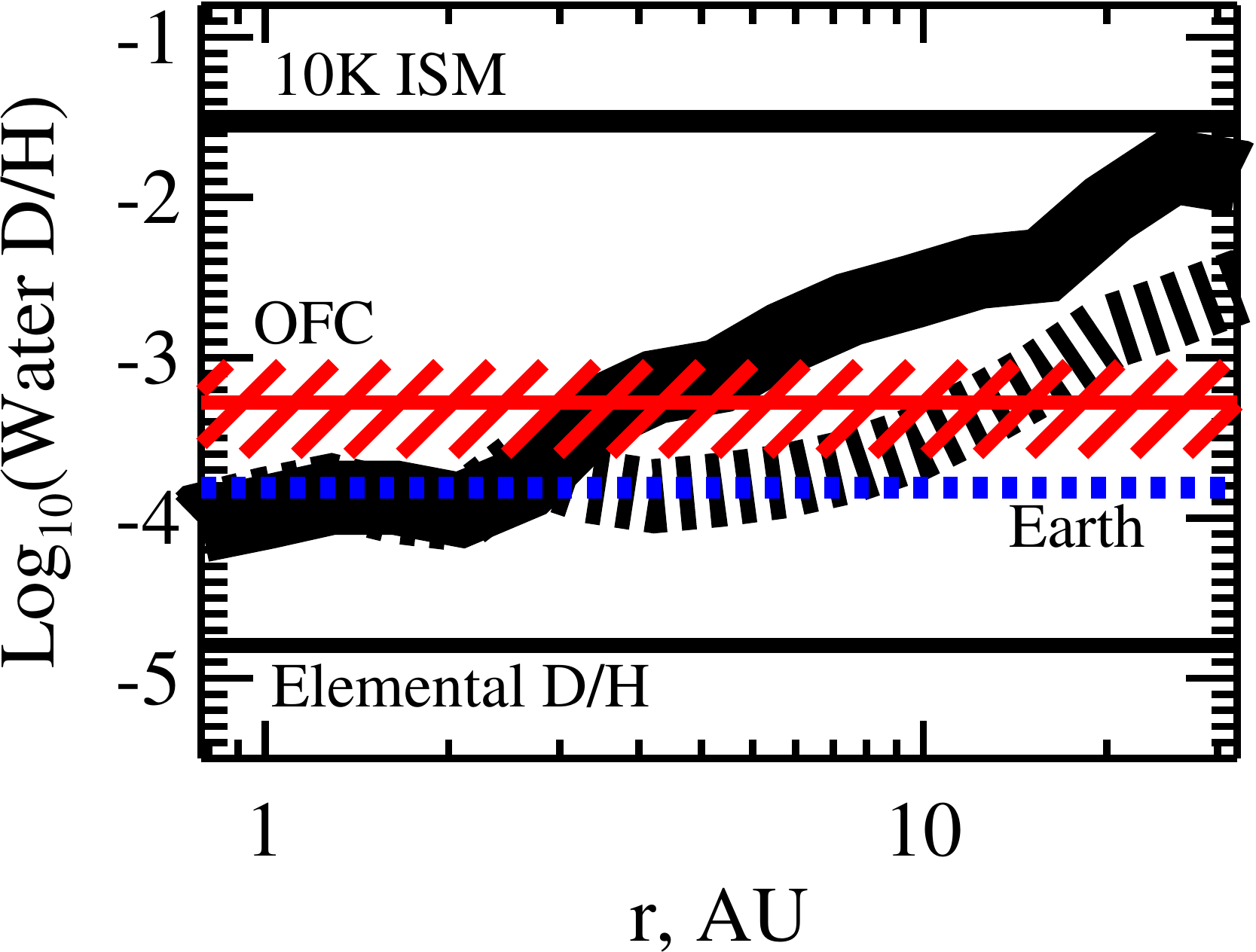} 
\caption{The radial distributions of the D/H ratios of the total water budget in the solar nebula between $1-30$~AU at 1~Myr are 
shown, both for the the laminar (solid line) and the 2D-mixing model (thick dashed line). The thickness of these lines reflects 
the uncertainties in the calculated water abundances, a factor of $\sim 2$~\citep{Vasyunin_ea08}. The elemental D/H ratio of 
$1.5\times10^{-5}$ is indicated by the thin straight solid line in the bottom \citep{1998ApJ...509....1S, 2003SSRv..106...49L}. 
The D/H ratio for water in the cold ISM, $3\times10^{-2}$, is depicted by the thick solid line on the top (our model). The Earth 
ocean's water D/H ratio, $1.59\times10^{-4}$ \citep{Lecuyer1998249}, is marked by the straight blue dashed line. The water D/H 
ratios in the Oort-family comets, which are a few times higher than the Earth value, are shown with the red rectangle filled with 
lines and denoted by ``OFC'' on the plot (see Table~\ref{tab:obs}).}
\label{fig:DH}
\end{figure}

The resulting distribution of the D/H ratio for the total water budget (ice + vapor) is shown in Figure~\ref{fig:DH}. The water 
D/H ratio of the Earth' oceans and the measured value in carbonaceous meteorites (``Earth") are shown for comparison, also similar 
to what observed in Jupiter-family comets. The D/H ratios for the Oort-family comets (``OFC'') are also indicated. 

As can be clearly seen, the laminar model shows the Earth water D/H ratio at $0.8-2.5$~AU, while for the 2D chemo-dynamical model 
such a low D/H value extends towards larger radii, $\la 9$~AU. Similarly, the elevated water D/H ratios representative of the 
Oort-family comets, $\sim 2.5-10\times 10^{-4}$, are achieved within $\sim 2-6$~AU and $\sim 2-20$~AU in the laminar and the 2D 
model, respectively. The reason for this behavior is a shallower radial gradient of the water D/H ratio in the 2D mixing model. 
Turbulent mixing slowly transports some of the water ice into warmer or irradiated regions where it desorbs and is quickly 
de-fractionated by ion-molecule and dissociative recombination processes in the gas phase. As a consequence, lower water D/H 
ratios can be retained further out in the nebula, which seems to be more consistent with the delivery of water on Earth by comets. 
Both the laminar and the mixing nebular models have the Earth' oceans and the Jupiter-family comets' water D/H ratios at radii 
$\sim 
2-3$~AU, where carbonaceous chondrites are believed to have formed. Thus both the laminar and dynamical nebular model can 
reproduce the water D/H ratios observed in carbonaceous asteroids. With regard to the water isotopic composition and the origin of 
the Jupiter-family and Oort-family comets, the mixing model seems to be favored over the laminar model. It allows for a larger 
region for their formation with the appropriate water D/H ratios, extending to the distance of $\sim 10-30$~AU. 

We have also studied how other processes may affect the results of our chemical modeling. First, given observational evidence for 
the presence of bigger grains in protoplanetary disks \citep[e.g.][]{Testi_ea03,2005A&A...437..189V, 2006A&A...446..211R, 
Bouwman_ea08,Perez_ea12}, for reviews see \citet{2008PhST..130a4014N, 2011ppcd.book..114H}, we have increased the uniform grain 
sizes to 1 and $10\mu$m and calculated the nebular water D/H ratios. The average $1\mu$m and $10\mu$m grain sizes imply 10 and 100 
times smaller surface area (per unit volume of gas), respectively, making gas-grain interactions less prominent, and allowing the 
high-energy radiation to penetrate deeper into the disk. The thermal desorption rates are independent of the grain size, while the 
cosmic-ray particle (CRP)-desorption rate depends on the grain size only weakly \citep{Leger_ea85}. The overall photodesorption 
rate increases because the nebula becomes more UV-transparent.

In contrast, the accretion rate on to the grains decreases when the total grain surface area per unit gas volume decreases. 
Consequently, in the inner warm nebular region at $r\la 10-20$~AU the amount of gaseous water increases by a factor of $\sim 10$ 
for $a_{\rm dust} = 1\mu$m and $\sim 100$ for $a_{\rm dust} = 10\mu$m, respectively, but it still constitutes only a tiny fraction 
of the total water budget. As a result, the water D/H ratio is not significantly affected and the three modeled D/H radial 
distributions are within the uncertainties of the calculated D/H values \citep[see][]{Vasyunin_ea08, 2013ApJS..207...27A}. The 
same is found for the nebular model with the standard grains but 100 times lower UV-desorption yield of $10^{-5}$.
 
Next, we studied how the calculated water D/H ratios are affected if only a limited set of the nuclear spin-state processes is 
included in our chemical network. For that test we calculated a model in which only the para states of H$_2$, H$_2^+$, and H$_3^+$ 
isotopologues were considered, implying more efficient deuterium isotope exchange. The water D/H ratios calculated with this model 
coincide with the ratios calculated with our standard model with ortho/para-species (within the factor of $\sim 3$ in intrinsic 
chemical uncertainties). 

This result is a combination of two extreme situations. First, the initial abundances are calculated using the physical conditions 
of a dark, dense, 10~K cloud core. At such conditions almost the entire H$_2$ population exists in the lowest energy para-state, 
and the isotope exchange reactions involving ortho-species are still out of equilibrium. This leads to similar values of the 
initial water D/H ratio. Second, in the water de-fractionation zone in the nebula, at $r \la 10-30$~AU, temperatures are higher 
than 30~K. These temperatures are high enough to enable rapid gas-phase de-fractionation regardless of the dominant nuclear 
spin-state form of H$_2$, H$_2^+$, and H$_3^+$, mainly due to X-ray- and CRP-driven reprocessing of CO by He$^+$, followed by 
rapid gas-phase production of H$_2$O. This combination of effects leads to the somewhat surprising result that a reduced network 
produce roughly the same abundances as the complete network. 

\begin{figure*}[!htb]
\centering
\includegraphics[width=0.99\textwidth]{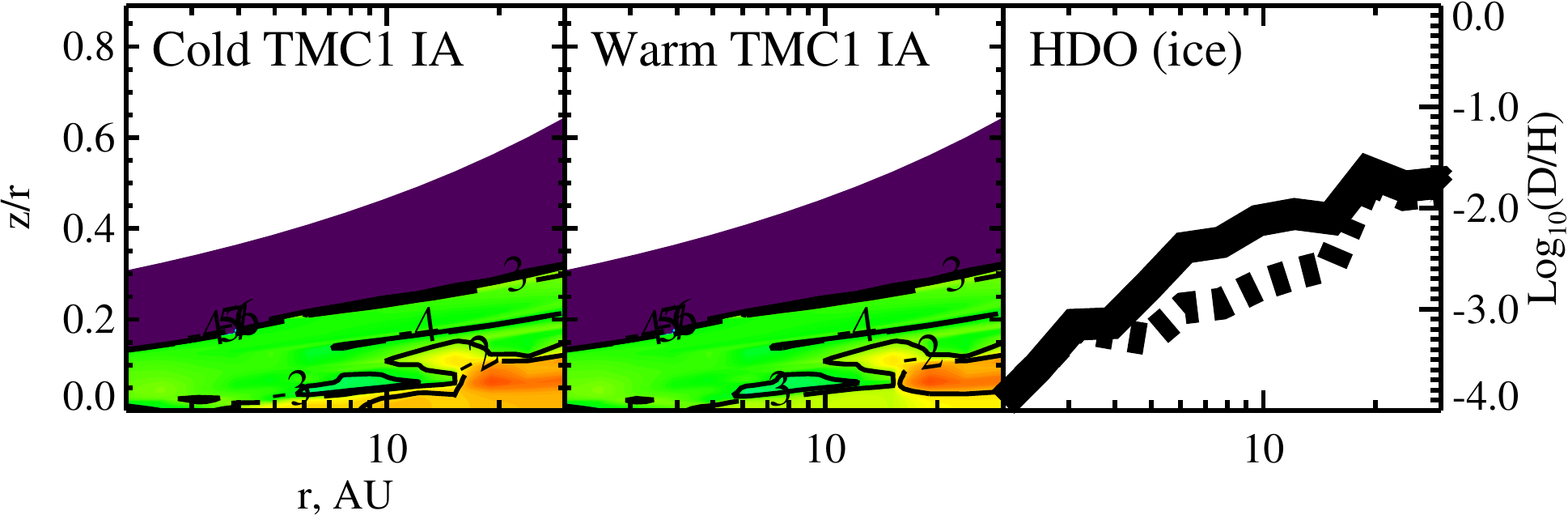} 
\caption{\bbf{The distributions of solid HDO D/H ratios in the solar nebula between 0.8 and 30~AU at 1~Myr. The `cold'' TMC-1 
initial abundance model is shown on the left panel and the ``warm'' model is shown in the middle panel. The vertically integrated 
column densities are compared in the right panel, with the ``cold'' TMC-1 initial abundance model depicted by solid line and 
``cold'' model depicted by dashed lines. The thickness of these lines correspond to intrinsic uncertainties in the calculated 
abundances and thus column densities.} }
\label{fig:gHDOopr}
\end{figure*}

\bbf{
The initial H$_{2}$ ortho:para ratio in cold dark clouds is unknown but predicted to be low \citep[$<0.1$][]{2009A&A...494..623P}. 
However, later, during the warmer protostellar phase this ratio is likely modified, such that there are more ortho$-$H$_2$ and 
less para$-$H$_2$ molecules. As such the ortho:para ratio can be closer to the statistical ortho:para ratio of 3:1 in the initial 
phase of the protoplanetary disk. We take the statistical value 3:1 as the upper extreme of the H$_{2}$ ortho:para ratio and 
initiate another run of the laminar model with initial abundances resulting from a TMC-1 model with an initial 3:1 ortho:para 
ratio for H$_{2}$. }

\bbf{In Figure~\ref{fig:gHDOopr} we show the D/H ratios of solid-phase HDO between 0.8 and 30 AU for the laminar model using our 
previous ``cold'' TMC-1 initial abundance (left panel) and ``warm'' TMC-1 initial abundance (middle panel), and in the right panel 
the D/H ratio determined from column densities is shown. In the disk midplane there are small but significant differences where 
D/H ratios decrease from $\sim 10^{-2}$ down to $\sim 10^{-4} - 10^{-3}$ around 10 AU. This cause the column density D/H ratios in 
the right panel to drop approximately an order of magnitude while the minimum and maximum D/H ratios remain unchanged at $\sim 1$ 
AU and $\gtrsim 30$ AU, respectively. Thus our conclusion that carbonaceous asteroids formed at 2-3 AU will inherit Earth 
water-like D/H ratios remains valid. The fact that the H$_{2}$O D/H ratio gets lower at intermediate radii ($\sim 10$ AU) will 
only help us to achieve a larger zone where Hartley-2-like comets with the Earth water D/H ratio could form, while still 
leaving a zone for the Oort comets to form at $\sim 10$ AU. 
}

Finally, we discuss the effects of the underlying chemistry on the D/H ratios throughout the disk.\bbf{In the midplane, 
temperatures are low enough to drive a rich surface chemistry, where O atoms are adsorbed on the grain surfaces and become 
hydrogenated, forming H$_{2}$O, and/or reacting with deuterium atoms to form HDO or D$_{2}$O. There is a small chance for the 
water molecules to be released into the gas-phase directly through chemisorption, but more likely they remain on the grains and 
desorb as material is transported into warmer regions, i.e. the inner regions or the atmosphere, where water, and its D/H ratio, 
will be reprocessed. 
}

The reprocessing of water occurs in the gas phase, so the relative adsorption and desorption plays an important role. These 
processes are heavily effected by temperature and density. In the warmer regions, the water abundances steadily increase by $\la 
30\%$ during the first 10\, 000 - 100\, 000 years. As soon as the water reaches the gas-phase, it is reprocessed. After that, the 
abundances of water ice and water vapor are at quasi steady-state. However, constant evaporation and re-accretion result in 
cycling of water between solid and gaseous phases. One important fact is that the D/H ratio never drops to the cosmic ratio, even 
in the warm, inner regions within a few AU of the Sun. In the inner solar nebula there are several important processes at work 
that sets the D/H ratio. First of all these regions experience strong X-ray or CRP-ionization and water is either ionized or 
protonated by reacting with various abundant ions, such as H$_{3}^{+}$, HCO$^{+}$, and their isotopologues:

\begin{eqnarray}
\small
\begin{tabular}{lclcc}
H$_{2}$O + $o-$H$_{3}^{+}$ 			$\Rightarrow$ H$_{3}$O$^{+}$	+	$o-$H$_{2}$; 		&	$4.5 
\times 10^{-9}$ cm$^{-3}$ s$^{-1}$		&	$\beta = -0.5$		\\
H$_{2}$O + $p-$H$_{3}^{+}$ 			$\Rightarrow$ H$_{3}$O$^{+}$	+	$o-$H$_{2}$; 		&	$2.3 
\times 10^{-9}$ cm$^{-3}$ s$^{-1}$		&	$\beta = -0.5$		\\
H$_{2}$O + $p-$H$_{3}^{+}$ 			$\Rightarrow$ H$_{3}$O$^{+}$	+	$p-$H$_{2}$; 		&	$2.3 
\times 10^{-9}$ cm$^{-3}$ s$^{-1}$		&	$\beta = -0.5$		\\
H$_{2}$O + HCO$^{+}$ 				$\Rightarrow$ H$_{3}$O$^{+}$	+	CO; 				&	
$2.1 \times 10^{-9}$ cm$^{-3}$ s$^{-1}$		&	$\beta = -0.5$		\nonumber
\end{tabular}
\label{eq:waterform}
\end{eqnarray}
\bbf{The same set of reactions applies for HDO being protonated into H$_2$DO$^+$ (as well as heavier isotopologues). The 
H$_{3}$O$^{+}$ isotopologues can then reform water through dissociative recombination, but the D/H ratios will be reset, 
characteristic of the temperature in the environment at which they are reforming:}
\begin{eqnarray}
\small
\begin{tabular}{lclcc}
H$_{3}$O + e$^{-}$ 			$\Rightarrow$ H$_{2}$O	+	H; 		&	$1.1 \times 10^{-7}$ cm$^{-3}$ 
s$^{-1}$		&	$\beta = -0.5$		\\
H$_{2}$DO + e$^{-}$ 		$\Rightarrow$ H$_{2}$O	+	D; 		&	$7.3 \times 10^{-8}$ cm$^{-3}$ s$^{-1}$		
&	$\beta = -0.5$		\\
H$_{2}$DO + e$^{-}$ 		$\Rightarrow$ HDO		+	H; 		&	$7.3 \times 10^{-8}$ cm$^{-3}$ 
s$^{-1}$		&	$\beta = -0.5$		\nonumber
\end{tabular}
\label{eq:waterform2}
\end{eqnarray}
The reprocessing of water involves several \bbf{additional} reactions that compete in both lowering and increasing the D/H ratio. 
In the warm inner nebular region the normal H$_{3}^{+}$~fractionation pathway is not active and instead fractionation is driven 
mostly via the ``warm'' fractionation route that includes light hydrocarbons such as CH$_2$D$^+$ and C$_2$HD$^+$ and is active at 
$T\la 80$~K \citep[e.g.,][]{1989ApJ...340..906M, 2005A&A...438..585R, 2009A&A...508..737P}. Consequently, the initially high D/H 
ratios of light hydrocarbons, $\sim 10^{-3}$, is not fully equilibrated with the cosmic D/H ratio of $\sim 10^{-5}$. 
Neutral-neutral reactions of these hydrocarbons with atomic or molecular oxygen are able to produce normal or heavy water, e.g.:
\begin{eqnarray}
\small
\begin{tabular}{lc}
CH$_{3}$ + O$_{2} \Rightarrow$ HCO + H$_{2}$O;	&	$1.66 \times 10^{-12}$ cm$^{-3}$ s$^{-1}$
\end{tabular}
\label{eq:warmfrac}
\end{eqnarray}
Another important channel of the water formation in the gas-phase is through the neutral-neutral reaction with OH or OD of 
H/D\bbf{ or $ortho/para-$H$_{2}$/HD}:
\begin{eqnarray}
\small
\begin{tabular}{lclcc}
OH + H 					$\Rightarrow$ H$_{2}$O;			&	$4.0 \times 10^{-18}$ cm$^{-3}$ s$^{-1}$	
	&			\\
OH + D 					$\Rightarrow$ HDO;				&	$4.0 \times 10^{-18}$ cm$^{-3}$ 
s$^{-1}$		&			\\
OD + H 					$\Rightarrow$ HDO;				&	$4.0 \times 10^{-18}$ cm$^{-3}$ 
s$^{-1}$		&			\\	
OH + $o/p-$H$_{2}$ 		$\Rightarrow$ H$_{2}$O 	+ H;		&	$8.4 \times 10^{-13}$ cm$^{-3}$ s$^{-1}$		
&	1040 K	\\
OH + HD 					$\Rightarrow$ HDO		+ H;		&	$4.2 \times 10^{-13}$ 
cm$^{-3}$ s$^{-1}$		&	1040 K	\\
OH + HD 					$\Rightarrow$ H$_{2}$O	+ D;		&	$4.2 \times 10^{-13}$ cm$^{-3}$ 
s$^{-1}$		&	1040 K	\nonumber
\end{tabular}
\label{eq:waterform3}
\end{eqnarray}
While OH/OD, $ortho/para-$H$_{2}$/HD and H/D are formed through various processes, including the dissociation of water, the D/H 
ratio of OH is dominantly controlled by the equilibrium of the following forward and backward processes:
\begin{eqnarray}
\small
\begin{tabular}{lclcc}
OH + D $\Rightarrow$	OD + H;	&	$1.3 \times 10^{-10}$ cm$^{-3}$ s$^{-1}$		&			
\label{eq:warmfrac1}\\
OD + H $\Rightarrow$	OH + H;	&	$1.3 \times 10^{-10}$ cm$^{-3}$ s$^{-1}$		&	810 K	
\label{eq:warmfrac2}
\end{tabular}
\end{eqnarray}
At around 100~K, the equilibrium state of this reaction will lead to an enhancement by a factor of approximately 100 in OD, which 
will transfer into the water D/H ratio, and competes with other de-fractionation processes, eventually leading to the enhanced D/H 
ratios of $\sim$ 10$^{-4}$. As we move further our in the disk the efficiency of reaction~\ref{eq:warmfrac2} gradually decreases, 
whereas the pace of water ice de-fractionation decreases due to slower cycling of water through gas and solid phases, and the 
water D/H ratio increases gradually until we reach the colder outer regions of the solar nebula. 

\bbf{Compared to \citet{2010MNRAS.407..232T}, we do not find that reactions formation of water through reactions between OH/OD and 
H$_{2}$/HD are the dominant neutral-neutral reactions, instead we find that OH/OD and H/D reactions are more dominant. This 
discrepancy is likely arising due to the energy barrier of reactions with H$_{2}$ or HD, while reactions with H or D are 
exothermic. Other differences between our models may also contribute to the discrepancies. \citet{2010MNRAS.407..232T} used a 
steady-state model to determine their key reactions, which are not exactly like ours. Bear in mind that time-dependent disk 
chemical models usually do not reach a chemical steady-state even after 1 Myr of evolution. If we would allow our calculations to 
run over much longer time scales, we could get a more similar set of key reactions as found by \citet{2010MNRAS.407..232T}. 
Furthermore, \citet{2010MNRAS.407..232T} adopt reaction rates from \citet{2007A&A...466.1197W} (H$_{2}$O) and 
\citet{1996CPL...253..177T} (
HDO), may also account for some of the discrepancy. This also result in different OH and OD column densities, where OH is 
approximately an order of magnitude more abundant in our model, decreasing by two order of magnitudes at $\sim 30$ AU, while OD is 
approximately an order of magnitude more abundant for \citet{2010MNRAS.407..232T}. Finally, the inclusion of ortho-para chemistry 
in our network, and differences in physical structures may contribute further to these differences. Most notably, the stellar 
parameters are signficantly different, whereas  \citet{2010MNRAS.407..232T} adopts values for a less massive (0.5 M$_\odot$), but 
larger (2 R$_\odot$ and cooler (4000 K) central protostar compared to our study. }

\section{Discussion}
We have shown the importance of 2D-mixing on the water D/H gradient in the young solar nebula, which pushes the (lower) water D/H 
ratios from the inner disk regions toward larger heliocentric distances as water is defractionated after being transported into 
the inner regions. Before discussing the implications of our results, it is important to discuss the limitations of our model. 

Foremost, protoplanetary disks are not static environments, but are ever dynamically evolving systems. Dust grains grow into 
planetesimals and eventually planets, with an associated evolution of physical conditions affecting the chemistry. As we 
previously discussed, a moderate increase of grain sizes had no significant effect on our results (see Section~\ref{sec:results}), 
but this might change with even larger particles. Variations in temperature and density will likely play a significant role. 
\citet{2011ApJ...741L..34C} predicted that significant variations in HDO/H$_{2}$O ratios can come from even small variations in 
dust temperatures at the time of ice formation. This would also affect initial abundances of our model, as the warm-up phase 
following the collapse of a dark pre-stellar core will experience a gradual increase in both temperature and density. In addition, 
exact masses and structural parameters of the solar nebula remain highly uncertain. 

Another large uncertainty is the post-processing of Earth's ocean water. Such processes include biological (life and its 
evolution) as well as chemical processes. Atmospheric processes can affect the fractionation if the young Earth had a massive 
hydrogen atmosphere, which would experience a slow hydrodynamic escape, which will be slower for deuterium compared to the lighter 
hydrogen isotope \citep{2008Icar..194...42G}. Furthermore, as was first proposed by \citet{2005...Campins}, processes involved in 
planetary accretion, such as outgassing and the evolution of hydrosphere and atmosphere, are complex and may have affected the 
fractionation. 

Finally, the giant impact theory which explains the formation of the moon and the similarities of its bulk composition to Earth 
\citep[see e.g.][]{2012Sci...338.1052C}, requires a collision between early Earth and a large impactor. This scenario implies a 
heavy loss of the primordial Earth atmosphere and would have likely greatly affected the chemical composition, more specifically 
the D/H ratio. However, results by \citet{Saal14062013} revealed a similar isotopic composition of the water in the interiors of 
Earth and the Moon, as well as the bulk water in carbonaceous chondrites, suggesting that this reprocessing is of minor 
importance. 

\begin{deluxetable}{lccc}
\centering
\tabletypesize{\small}
\tablecaption{Compilation of observed deuterated water in disks and comets.\label{tab:obs}}
\tablehead{
\colhead{Source}				&	\colhead{D/H ratio $[\times$ $10^{-4}]$}	&	
\colhead{References}		\\}
\startdata
solar nebula					&		$1.7-2.5$						&	1	
		\\
Standard Mean Ocean Water		&		$1.46 -1.52$ 					&	2			
\\
Carbonaceous chondrites		&		$1.30-1.50$					&	3			
\\[0.07cm]
\multicolumn{3}{l}{Protoplanetary disks}\\ 		\hline\\[-1.5ex]
LkCa 15						&		640							&	4	
		\\
DM Tau						&		$\gtrsim 100$ 					&	5		
	\\[0.07cm]
\multicolumn{3}{l}{Oort-family comets}\\ 				\hline\\[-1.5ex]
Halley						&	2.55$-$3.46						&	6		
	\\
Halley						&	2.72$-$3.40						&	7		
	\\
Hyakutake					&	1.90$-$3.90						&	8		
	\\
Hale-Bopp					&	2.50$-$4.10						&	9		
	\\
Ikeya-Zhang					&	$<$ 3.1							&	10		
	\\
C/2002 T7					&	1.80$-$3.20						&	11		
	\\ 
C/2001 Q4					&	3.2$-$6.0							&	12	
		\\  
8p/Tuttle						&	2.64$-$5.54						&	13	
		\\
Machholz						&	2.71$-$3.69						&	14	
		\\
C/2007 W1 (Boattini)			&	$< 12.9$							&	15		
	\\
C/2007 N3 (Lulin)				&	$<$ 5.6							&	16		
	\\
C/2009 P1 (Garradd)			&	1.84$-$2.28						&	17			
\\[0.07cm]
\multicolumn{3}{l}{Jupiter-family comets}\\				\hline\\[-1.5ex]
103P/Hartley 2					&	1.37$-$1.85						&	18		
	\\H
45P/Honda-Mrkos-Pajdusakova	&	$<$ 2							&	19			
\enddata
\tablerefs{(1) \citet{2001A&A...370..610L};
(2) \citet{Lecuyer1998249}; 
(3) Compiled from \citet{Boato1954209, Robert198281, Kerridge19851707};
(4) \citet{2003cdsf.conf..188K};
(5) \citet{2005ApJ...631L..81C}; 
(6) \citet{1995JGR...100.5827B};
(7) \citet{1995A&A...302..301E};
(8) \citet{1998Icar..133..147B};
(9) \citet{1998Sci...279..842M};
(10) \citet{2006A&A...449.1255B};
(11) \citet{2008A&A...490L..31H};
(12) \citet{2008LPICo1405.8216W}; 
(13) \citet{2009ApJ...690L...5V};
(14) \citet{2009ApJ...693..388K};
(15) \citet{2011Icar..216..227V};
(16) \citet{2012ApJ...750..102G};
(17) \citet{2012A&A...544L..15B};
(18) \citet{2011Natur.478..218H};
(19) \citet{2013ApJ...774L...3L}
}
\end{deluxetable}

\subsection{Observations in other protoplanetary disks}
\citet{2013ApJ...766..134N} found a clear relation between the mid-infrared HCN/H$_{2}$O flux ratio and submillimeter disk mass 
among T Tauri stars in Taurus. They argue that this interesting trend arises as a consequence of a more efficient formation of 
large non-migrating bodies in more massive disks that lock up oxygen and water beyond the water line. This depletion of oxygen 
enhances the gas C/O ratio which results in an increase of HCN and decrease of H$_{2}$O abundances. Thus, we see this dramatic 
trend in the HCN/H$_{2}$O ratios with disk mass. \citet{2011ApJ...733..102C} found that only modest enhancements in the C/O ratio 
in inner disk atmospheres are needed for a significant increase of warm HCN/H$_{2}$O. This efficient lock-up of water in massive 
disks is necessary to bear in mind when comparing to observations of other protoplanetary disks, as their masses may differ from 
that of the young solar nebula ($\sim 0.11$ M$_{\odot}$).
\bbf{TW Hydrae has been observed on multiple occasions, and its expected} disk mass is comparable to the solar nebula and believed 
to be able to form a planetary system like our own \citep{2013Natur.493..644B}, hence one may expect a similar chemistry. 
\citet{2008ApJ...681.1396Q} estimated water column densities of $<~9.0~\times~10^{14}$ cm$^{-2}$ in the outer region of TW Hydrae, 
which is lower compared to our models ($\sim 10^{15}$ cm$^{-2}$). \bbf{A small difference in disk mass may, partly, explain the 
discrepancy, as has been suggested by \citet{2013ApJ...766..134N} from studies of HCN/H$_{2}$O flux ratios, where more massive 
disks have a more efficient formation of large non-migrating bodies, which more efficiently locks up oxygen and water beyond the 
water line.} \citet{2011Sci...334..338H} observed a cold water vapor reservoir in the disk of TW Hydrae, predicting instead a 
relative abundance $\sim 10^{-7}$ at intermediate heights, slightly higher than we predict in the upper layers of the solar 
nebula ($\sim 10^{-8}$). It is possible that they probe regions further in than \citet{2008ApJ...681.1396Q}, due to the optical 
depth-effect, corresponding to ``warmer'' water at $\sim 50$ AU. 

\citet{2005ApJ...622..463P} observed water ice features in an edge-on circumstellar disk using the \textit{Spitzer} telescope. The 
orientation means that their observations probe the outermost regions. The high inclination complicates comparisons and the cutoff 
radius of the disk will affect results. Using models and adopting the best-fitting parameters to their observations, 
\citet{2005ApJ...622..463P} estimated an abundance of $\sim 10^{-4}$ with respect to H$_{2}$, which is in agreement with our 
modeled relative abundances in the outer disk, $10^{-5} - 10^{-4}$. \citet{2012A&A...538A..57A} measured with the AKARI satellite 
several ice features in five edge-on Class II disks with column densities on the order of $10^{17}$ cm$^{-2}$ as well as a faint 
HDO feature in one of the disks (HV Tau, D/H $\sim 19\%$). The observed disk has a high inclination ($i \sim 84^{\circ}$) and 
probe the outermost regions where our models estimate a water D/H ratio of $\sim 0.03-0.05$, a D/H ratio lower by a factor $\sim 
5$ 
compared to the observed ice features. Our column densities are vertically integrated, and due to the higher density in the disk 
midplane compared to the surface, horizontally integrated column densities would be expected to be larger by approximately one 
order of magnitude. 

\subsection{Previous theoretical studies}
There has been a vast number of studies on the chemistry in protoplanetary disks, many of these are not targeting the problem in a 
similar fashion as us, making any comparison complicated. Therefore, we concentrate on studies with similar approaches to model 
chemistry and physics in a protoplanetary disk or solar nebula.

\citet{1999ApJ...526..314A} studied the deuterium chemistry in the outer regions of protoplanetary disks with an 1D accretion 
flow, using a collapse model to set up the initial molecular concentrations. They have found that the molecular D/H ratios are 
enhanced with respect to the protosolar values, and that the ratios at $\sim 30$~AU agree reasonably well with the D/H ratios 
observed in comets, meaning that comets may not necessarily be composed of primordial, unprocessed interstellar matter, \bbf{as we 
also find in our models. While our two models predict similar abundances, within an order of magnitude, we find different behavior 
of D/H ratios with radii, such that our models attain higher D/H ratios in the outer regions. } 

\cite{2007ApJ...660..441W} have investigated deuterium chemistry in outer disk regions, using the UMIST RATE'95 database extended 
with a set of reactions for multiply-deuterated species, a 1+1D disk model of \cite{2001ApJ...553..321D}, and initial molecular 
abundances obtained by the chemical modeling of a cold prestellar core. Furthermore, they implement a lower stellar and disk mass 
compared to our model. Similarly to \cite{1999ApJ...526..314A}, they found that the D/H ratios observed in comets may partly 
originate from the parental molecular cloud and partly be produced in the disk. They concluded that the D/H ratios of gaseous 
species are more sensitive to deuterium fractionation processes in disks due to rapid ion-molecule chemistry compared to the 
deuterated ices, whose D/H values are regulated by slow surface chemistry and are imprints of the cold conditions of the 
prestellar cloud. \bbf{We find signs of the same sensitivities to ion-molecule chemistry, which is, in our models, also further 
aided 
by high-temperature reactions.}

In their later study, \cite{2009ApJ...703..479W} investigated deuterium chemistry in the inner 30~AU, accounting for gas and dust 
thermal balance. While a good agreement between the model predictions and observations of several non-deuterated gaseous species 
in a number of protoplanetary disks was obtained, the calculated D/H ratios for ices were higher than measured in the Solar System 
Oort-family comets, \bbf{while our models are in better agreement to observations of Solar System bodies (see 
Section~\ref{sec:SS}). Furthermore, we note that their D/H distributions do not show a smooth increase with radii, as our models 
predict, but which show a spike and swinging variations at radii $\gtrsim 10$ AU.} \cite{2009ApJ...703..479W}, however, consider a 
small stellar mass and less turbulent mixing ($\alpha = 0.025$). Their results point to the importance of dynamical processes 
(shocks, turbulent or advective mixing, non-steady accretion) for deuterium chemistry in the inner regions of disks. 

\citet{2010MNRAS.407..232T} focused on understanding deuterium fractionation in the inner warm disk regions and investigated the 
origin of the high H$_2$O D/H ratios in dense ($\ga 10^6$~cm$^{-3}$) and warm gas ($\sim 100-1000$~K) by gas-phase photochemistry 
(dominated by photoprocesses and neutral-neutral reactions). Using the time-dependent chemical model based the UMIST RATE'06 
database~\citep{2007A&A...466.1197W} and the T~Tau disk structure calculated with ``ProDiMo'' \citep{2009A&A...501..383W}, they 
predicted that in the terrestrial planet-forming region at $\la 3$~AU the water D/H ratios may remain high, $\ga 1\%$, which is 
significantly higher than the value of $\approx 1.5\times 10^{-4}$ measured in the Earth ocean water as well as predictions from 
our models. While \citet{2010MNRAS.407..232T} a dynamical code, they only included a simple chemical network and did not account 
for the grain chemistry, which is essential for the calculation of water chemistry. 

\citet{Yang2013} coupled a classical dynamic $\alpha$-viscosity model of material transport, which calculates the evolution of the 
disk's surface density profile, and mixing with a kinetic study of D-H isotopic exchange amongst gas-phase molecules. They found 
that the water D/H ratio is low in the hot inner disk due to rapid exchange reactions with molecular hydrogen and increases 
outwards where the exchange becomes less efficient. Contrary to previous studies, they found that further out in the disk, the 
water D/H ratio decreases again as water exchanged at high temperatures near the young star is transported outwards in the early 
evolutionary stages of the disk. However, their chemical approach is simplistic and only includes neutral-neutral reactions as 
these would be dominant in the early stages of disk evolution because effects of ion-molecule reactions and photochemistry are 
significantly diluted. With the two-dimensional mixing in effect, we however find that these processes play a role in the 
chemistry and deuterium fractionation. 

\bbf{Recently, \citet{2013arXiv1310.3342F} investigated the water chemistry in turbulent protoplanetary disks. They found that 
transport by turbulence in to the atmosphere allows a reprocessing of water, which is destroyed by photoreactions and then 
transported back to the midplane where water is (re)formed, and this cycle is most effective at radii $\lesssim 30$ AU. While 
\citet{2013arXiv1310.3342F} also discussed the effects of radial transport without including it in their models, we have done so 
and found that radial transport can further decrease both water abundances and D/H ratios through reprocessing in the warmer, 
inner midplane regions. They also studied the effects from different desorption energies for atomic hydrogen and deuterium, 
comparing $E_{des, H/D} = 400$ K and $E_{des, H/D} = 600$ K \citep[as found from molecular dynamics simulations for crystalline 
and amorphous water ice, respectively
]
{2007MNRAS.382.1648A}. The higher energy increases the residence time on grains for H and D, and results in a higher water reformation rate in the midplane, and hence the resulting water abundances are higher by up to an order of magnitude.. The most significant difference to our model is in the calculated D/H ratios throughout the disk. For \citet{2013arXiv1310.3342F}, in the inner region, D/H drops to the cosmic ratio $\sim 1.5\times 10^{-5}$ at radii $\lesssim 10$ AU, and even lower at radii $\lesssim 2$ AU, while our 2D-mixing model retain an enhanced D/H ratio of $\sim 10^{-4}$ between $\sim 1-7$ AU. We find that the enhanced ratio at radii $\lesssim 7$ AU is largely because of ``warm'' fractionation pathways (see Section~\ref{sec:results}), which is included in both models. It is therefore likely that this discrepancy is a result of different physical parameters, such as stellar parameters adopted, leading to different thermal structure of the nebula. 
}

\subsection{Solar System bodies}\label{sec:SS}
There is a great variation in chemical composition in the different bodies of the Solar System. Due to the temperature gradient throughout the solar nebula, \citep[see figures in][]{2011Natur.478..218H,2013ApJ...774L...3L}, the D/H ratio varies radially through the disk, as we have also seen in our results (see Figure~\ref{fig:DH}). The same behavior has been observed in the solar nebula from observations of small Solar System bodies \citep{2000SSRv...92..201R}: the D/H ratios increasing with radial distance from the Sun. Due to sedimentation in the planetary cores and reprocessing in their atmospheres, the D/H ratios in planetary bodies of the Solar System are no direct measure of the pristine D/H ratio.

While we can understand the rough composition of most of the planets in the Solar System, the more important discussion is the composition of small Solar-System bodies, i.e. asteroids and comets. Their origin is less constrained, but we can derive boundary conditions from the Grand Tack scenario \citep{2012AREPS..40..251M}, the most successful theory of the formation and evolution of the Solar System. In this scenario, Jupiter migrated inwards after its formation, only to be stopped at $\sim 1.5$ AU by a mean motion resonance with Saturn after which both giant planets move back outwards to their present positions, and before Jupiter had managed to diminish the available material in the inner disk for build-up of the terrestrial planets. In this theory, many of the observed features of the Solar System today can be reproduced, such as the smaller mass of Mars \citep{2011Natur.475..206W}. As Jupiter moved through the asteroid belt at $\sim$2-3 AU twice, it scattered much of the belt material. This might 
explain the mix of C- and S-type asteroids in the inner and outer asteroid belts \citep{2012M&PS...47.1941W}. 

Carbonaceous chondrites have measured water D/H ratios very similar to that of Earth's oceans \citep{2003SSRv..106...87R}. Comets, or more specifically Oort-family comets, on the other hand, have been observed with enhanced D/H ratios relative to the Earth's oceans. Observations reveal values $\sim $1.8-6.0~$\times$~10$^{-4}$ (see Table~\ref{tab:obs}), an enhancement to the Earth ocean value by a factor of few. These values would put their formation origin in our models around $10-20$ AU. However, new experimental studies of ice sublimation suggest that the D/H measured in the evaporated vapor of comets might be depleted by 70\% or more with respect to the bulk D/H ratio in the nucleus \citep{2012P&SS...60..166B}. With this in mind, the measurements of D/H ratios in Oort-family comets would be even more enhanced relative to Earth's oceans, by more than an order of magnitude. Regarding the origin of these comets, it means that they would be expected to have originated further out in the disk, $\sim 30-40$ AU, 
and is in agreement that these long-period comets are believed to have formed much further in compared to where they are found today in the Oort cloud \citep[see e.g.][]{1950BAN....11...91O, 1987AJ.....94.1330D}.

While Oort-family comets have too high D/H ratios to be considered as a major source in the delivery of Earth's ocean, the observation by \citet{2011Natur.478..218H} has revealed D/H ratios similar to Earth's ocean water in the Jupiter-family comet 103P/Hartley-2. The existence of diversity in D/H ratios between the Jupiter- and Oort-family comets has been further confirmed by the observations of \citet{2013ApJ...774L...3L} who observed a D/H ratio $< 2 \times 10^{-4}$ in the Jupiter-family comet 45P/Honda-Mrkos-Pajdusakova. Therefore, we may consider the D/H ratio of Hartley-2 as representative of Jupiter-family comets, which puts their possible formation location between $\sim$ 1-10 AU. Most likely they originate in the region between Jupiter and Saturn where the gravitational pull of Jupiter and Saturn have scattered them, giving them their current high-eccentricity orbits past the snow line $\sim 3$ AU. If they would have formed in the Kuiper belt located beyond the orbit of Neptune, which is the current 
theory \citep[see e.g.][]{1999SSRv...90..301W, 2007MNRAS.381..779E, 0004-637X-687-1-714}, their D/H ratios would have been much higher in our model as the chemical timescale for de-fractionation of HDO ice is too long. It still means that the origins of the Jupiter- and Oort-family comets are not as distinct as previously thought. 

\section{Conclusions}
In this paper the isotopic and chemical evolution of water in the early history of the solar nebula before the onset of planetesimal formation is studied. An extended gas-grain chemical model that includes multiply-deuterated species, high-temperature reactions, and nuclear spin-state processes is combined with a 1+1D steady-state $\alpha$-viscosity nebula model. To calculate initial abundances, we simulated the 1~Myr of the evolution of a cold and dark TMC1-like prestellar core, resulting in initially high D/H ratios for water and other molecules of $\sim 1\%$. Two time-dependent chemical models of the solar nebula are calculated over 1~Myr and for radii $0.8-800$~AU: (1) a laminar model and (2) a model with 2D-turbulent mixing transport. We find that both models are able to reproduce the Earth ocean's water D/H ratio of $\approx 1.5\times 10^{-4}$ at the location of the asteroid belt, $\la 2.5-3$~AU, where a transition from predominantly solid to gaseous water occurs.

The water ices there can be incorporated in growing solids, melt, and eventually produce phyllosilicates. At $\la 2$~AU, nebular temperatures are too high for the water ice to exist and the dust grains are water ice-free. Thus the planetesimals, from which Earth would later form, remain water-poor. We find that the radial increase of the D/H ratio in water outward is shallower in the chemo-dynamical nebular model. This is related to more efficient de-fractionation of HDO via rapid gas-phase processes, as the 2D mixing model allows the water ice to be transported either inward and thermally evaporated or upward and photodesorbed. Taking the water D/H abundance uncertainties of the factor of 2 into account, the laminar model shows the Earth water D/H ratio at $r \la 2.5$~AU, while for the 2D chemo-dynamical model this zone is larger, $r \la 9$~AU. Similarly, the enhanced water D/H ratios representative of the Oort-family comets, $\sim 2.5-10\times 10^{-4}$, are achieved within $\sim 2-6$~AU and $\sim 6-30$~AU 
in the laminar and the 2D model, respectively. The characteristically, slightly lower, water D/H ratio that has been found for Jupiter-family comets are found further in and we find their possible formation location $\sim 1 - 10$ AU in both models. This means that, in our models, we find an overlap in the possible formation location for Oort- and Jupiter-family comets. However, with regard to the water isotopic composition and the origin of the comets, the mixing model seems to be favored over the laminar model as the former allows Oort-family comets to have formed in the region of Jupiter's and Saturn's present location. 

\acknowledgments
This research made use of NASA's Astrophysics Data System. The research leading to these results has received funding from the European Community's Seventh Framework Programme [FP7/2007-2013] under grant agreement no. 238258. DS acknowledges support by the {\it Deutsche Forschungsgemeinschaft} through SPP~1385: ``The first ten million years of the Solar System - a planetary materials approach'' (SE 1962/1-2).

\bibliographystyle{apj}
\bibliography{DeuteratedDiskWater}

\end{document}